\documentclass[amsmath,amssymb,superscriptaddress,twocolumn]{revtex4-2}

\usepackage{graphicx}
\usepackage{dcolumn}

\def\be{\begin{equation}}
\def\ee{\end{equation}}
\def\beq{\begin{eqnarray}}
\def\eeq{\end{eqnarray}}

\usepackage{bm}
\usepackage{graphicx}
\usepackage{dcolumn}
\usepackage{amsmath}
\usepackage[latin1]{inputenc}
\usepackage{graphicx, psfrag}
\usepackage{amssymb}
\usepackage[colorlinks=true, citecolor=blue, urlcolor = blue, linkcolor= red, bookmarks=true]{hyperref}
\usepackage{float}
\usepackage{amsmath}
\usepackage{amsfonts}
\usepackage{dcolumn}
\usepackage{hyperref}
\usepackage{subfigure}
\usepackage{pgfplots}
\usepackage{epstopdf}
\usepackage{booktabs}

\begin{document}


\title{Wormhole geometries in $f(Q)$ gravity and the energy conditions}



\author{Ayan Banerjee} \email[]{ayanbanerjeemath@gmail.com}
\affiliation{Atrophysics Research Centre, School of Mathematics, Statistics and Computer Science, University of KwaZulu--Natal, Private Bag X54001, Durban 4000, South Africa}

\author{Anirudh Pradhan} \email[]{pradhan.anirudh@gmail.com}
\affiliation{Department of Mathematics, Institute of Applied Sciences and Humanities, GLA University, Mathura-281 406, Uttar Pradesh, India}

\author{Takol Tangphati} 
 \email[]{takoltang@gmail.com}
\affiliation{Department of Physics, Faculty of Science, Chulalongkorn University, \\Bangkok 10330, Thailand}

\author{Farook Rahaman}%
\email[Email:]{rahaman@associates.iucaa.in}
\affiliation{Department of Mathematics, Jadavpur University, Kolkata-700032, India}


\date{\today}

\begin{abstract}
Following the recent theory of $f(Q)$ gravity, we continue to investigate the possible existence of wormhole geometries, where $Q$ is the non-metricity scalar. Recently, the non-metricity
scalar and the corresponding  field equations have  been  studied for some spherically symmetric configurations in [ Phys. Lett. B 821, 136612 (2021) and  Phys. Rev. D 103,  124001 (2021) ]. One can note that field equations are different in these two studies. Following [Phys. Rev. D 103, 124001 (2021)], we systematically study the field equations for wormhole solutions and found the violation of null energy conditions in the throat neighborhood. More specifically, considering specific choices for the $f(Q)$ form and for constant redshift with different shape functions, we present a class of solutions for static and spherically symmetric wormholes. Our survey indicates that wormhole solutions could not exist for specific form function $f(Q)= Q+ \alpha Q^2$. To summarize, exact wormhole models can be constructed with violation of the null energy condition throughout the spacetime while being $\rho \geq 0$ and vice versa. 

\end{abstract}


\maketitle


\section{Introduction}\label{intro:Sec}


The idea of wormholes act as tunnel-like structures that connect two parallel universes or distant parts of the same universe.  It was J.A. Wheeler \cite{Fuller1957} who first introduce
the term wormhole as objects of the spacetime quantum foam connecting different regions of spacetime at the Planck scale. Although these solutions were not traversable and collapsed instantly upon formation, as insightfully reviewed in \cite{Fuller1962}. Modern interest in wormhole physics was stimulated after the seminal work of Morris and Thorne in 1988 \cite{Morris:1988cz}. They considered static and spherically symmetric line elements and  discussed the mechanism for traversable wormholes. The traversability assumes that matter and radiation can travel freely in both directions and in a reasonable time through the wormhole. Subsequently,  Morris, Thorne and Yurtsever \cite{Morris1988} came up with an idea that wormhole can be converted
into a time machine with which causality might be violated.   For more information we refer the reader to the vast literature on wormholes, see Refs. \cite{Visser:1995,Lobo:2007zb}.

However, it is well known in general relativity that wormhole spacetimes are supported by exotic matter whose stress-energy tensor violates the null energy condition (NEC) \cite{Morris:1988cz,Visser:1995}, according to the needs of the geometrical structure. In fact, traversable wormholes violate all of the pointwise energy conditions and averaged energy conditions \cite{Lobo:2007zb}. However, in \cite{Konoplya:2021hsm} authors have found solutions describing asymmetric asymptotically flat traversable wormholes supported by ordinary Dirac and Maxwell fields.  
Since, the exotic matter is a problematic issue and thus many arguments have been given in favor of the violation of the energy conditions such as invoke quantum fields in curved spacetime,  scalar-tensor theories and so on.
So many attempts have been made to minimize the use of exotic matter. Among them ``volume integral quantifier" is one of the most popular approaches which quantifies the total amount of energy condition violating matter \cite{Visser:2003yf,Kar:2004hc}. This formulation was
further improved by  Nandi \emph{et al} \cite {Nandi:2004ku} to know the \textit{exact quantity} of exotic matter present in a given spacetime.  Further, there have been proposals regarding confinement
of exotic matter at the throat of the wormhole, namely, the cut and paste procedure  see Refs. \cite{Visser:1989kh,Visser:1989kg,Lobo:2003xd} for more. According to this process, interior solution is being matched with an exterior vacuum solution at a junction interface, where the wormhole throat is located.  During the past decades, there have been a lot of 
research exploring the possible existence of wormhole geometries supported by the
exotic equation of state (EoS) \cite{Lobo:2005us,Sushkov:2005kj}, and further developed in \cite{Jamil:2010ziq,Zaslavskii:2005fs,Bronnikov:2006pt,Gonzalez:2009cy,Cataldo:2008ku}).

It is an accepted fact that constructing a wormhole with ordinary matter (i.e., satisfy the energy conditions) has been a big challenge in gravitation physics.  It was  shown that higher-dimensional cosmological wormholes \cite{Zangeneh:2014noa} and wormholes in modified theories of gravity 
\cite{Mehdizadeh:2015jra,Banerjee:2020uyi,Mazharimousavi:2016npo,Bronnikov:2002rn} can be constructed without 
exotic matter, at least in the neighbourhood of the throat. In fact, in the context of $f(R)$ theories of gravity, the solution of wormholes have attracted much attention where 
wormholes can be theoretically constructed with the presence of normal matter  \cite{Pavlovic:2014gba,Lobo:2009ip}. This type of solutions were also found in  $f(T)$ gravity \cite{Boehmer:2012uyw,Sharif:2013exa}, hybrid metric-Palatini theory \cite{Capozziello:2012hr}, multimetric gravity \cite{Hohmann:2013dra}, Rastall gravity \cite{Moradpour:2016ubd}, conformal Weyl gravity \cite{Lobo:2008zu,Varieschi:2015wwa}, modified gravity \cite{Harko:2013yb}, Horndeski theory of gravity \cite{Korolev:2020ohi} and other theories.

In this article our main interest is to explore the possible existence of wormhole solutions in a recently developed symmetric teleparallel (ST) gravity or $f(Q)$ gravity theory,  where $Q$ is the non-metricity scalar \cite{BeltranJimenez:2017tkd}. The key difference between ST and GR is the role played by the affine connection,  $\Gamma^{\alpha}_{\mu\nu}$ rather than the physical manifold. Most remarkably, $f(Q)$ gravity
is equivalent to GR in flat space \cite{BeltranJimenez:2017tkd}. It is important to keep in mind that similar to the $f(T)$ gravity, $f(Q)$ gravity also features in second order field equations,  while gravitational field equations of $f(R)$ gravity are of the fourth-order \cite{Sotiriou:2008rp}.  Thus, $f(Q)$ gravity provides a different geometric description of gravity, which is nevertheless equivalent to GR. In \cite{DAmbrosio:2021zpm}, authors have systematically derived and studied symmetry reduced field equations for $f(Q)$ gravity. 
Along with the increasing interest on $f(Q)$ gravity, several solutions have been widely studied  in the cosmological setting, see e.g., 
Refs. \cite{Lazkoz:2019sjl,Barros:2020bgg,Mandal:2020buf,Ayuso:2020dcu,Frusciante:2021sio,Khyllep:2021pcu,Esposito:2021ect,BeltranJimenez:2019tme}. 

However,  in such a theory only few solutions have been found in static and spherical symmetric spacetime. Spherically symmetric configuration in $f(Q)$ gravity was considered in \cite{Lin:2021uqa}, and explored the application of this theory considering stellar structure with polytropic equation of state (EoS). In a recent study,  wormhole solutions from the Karmarkar condition have been obtained and studied  in $f(Q)$ gravity extensively \cite{Mustafa:2021ykn}. In the present manuscript 
our interest is to find exact and correct field equations in $f(Q)$ gravity for static and spherical symmetric configuration. We further extend this analysis and find an exact wormhole solution, where we showed the violation of the NEC of normal matter at the throat of the wormhole. 

The present paper is organized as follows: In Section \ref{secII} we give an overview about the $f(Q)$ gravity, and then we find the corresponding field equations for static and spherically symmetric spacetime in  Section \ref{sec3}. In the same section, we find exact solutions
of wormhole geometries in $f(Q)$ gravity, paying close attention to the energy conditions and outlining different approaches in finding specific solutions. Finally we give our conclusions in section \ref{sec8}.

\section{Setting the stage: $f(Q)$ gravity}
\label{secII}

In the present work, we consider the action for $f(Q)$ gravity \cite{BeltranJimenez:2017tkd} is given by
\be  \label{qqm}
 S=\int  \left[\frac{f(Q)}{16\pi} +\mathcal{L}_m\right]\sqrt{-g}~d^4x,
 \ee
where $f(Q)$ is an arbitrary function of the non-metricity $Q$, $g$ is the determinant of the metric $g_{\mu\nu}$ and ${\cal L}_m$ is the Lagrangian density corresponding to matter. We define the non-metricity tensor by
\begin{equation}
    \label{Qcom}
    Q_{\alpha\mu\nu}=\nabla_\alpha g_{\mu\nu}=-L^\rho_{\alpha\mu}g_{\rho\nu}-L^\rho_{\alpha\nu}g_{\rho\mu},
\end{equation}
where the term disformation is given by 
\begin{equation}
    \label{Qcom1}
    L^\alpha_{\mu\nu}=\frac12Q^\alpha_{\mu\nu}-Q_{(\mu\nu)}^{\quad\;\alpha},
\end{equation}
and the two independent traces of the non-metricity tensor are as follows:
\begin{equation}
Q_\alpha = Q_{\alpha }{}^{\mu }{}_{\mu  } \,,\quad \tilde{Q}_\alpha = Q^{\mu }{}_{\alpha \mu  }\,.
\end{equation}%
In this case the non-metricity scalar is defined as a contraction of $Q_{\alpha\beta\gamma}$ which is given by 
\begin{equation}
    \label{Qdef1}
    \begin{split}
        Q=&-g^{\mu\nu}\left( L^\alpha_{\beta\nu}L^\beta_{\mu\alpha}-L^\beta_{\alpha\beta}L^\alpha_{\mu\nu} \right)\\
        =&-P^{\alpha\mu\nu}Q_{\alpha\mu\nu}.
    \end{split}
\end{equation}
where $P^{\alpha \beta \gamma}$  is the non-metricity conjugate 
and the corresponding tensor is written as
\begin{eqnarray}
4P^{\alpha }{}_{\mu\nu } &=& -Q^{\alpha }{}_{\mu\nu} + 2Q_{(\mu %
\phantom{\alpha}\nu )}^{\phantom{\mu}\alpha } - Q^{\alpha }g_{\mu\nu }  \notag
\\
&&-\tilde{Q}^{\alpha }g_{\mu\nu}-\delta _{(\mu }^{\alpha }Q_{\nu )}\,. \label{super}
\end{eqnarray}
Now, the variation of (\ref{qqm}) with respect to $g_{\mu\nu}$ gives the field equations 
\begin{eqnarray}
&&\frac{2}{\sqrt{-g}}\nabla_\alpha\left(\sqrt{-g} f_Q P^\alpha{}_{\mu\nu}\right)
+ \frac{1}{2}g_{\mu\nu} f  \notag \\
&& + f_Q \left( P_{\mu\alpha\beta}Q_{\nu}{}^{\alpha\beta}
-2Q_{\alpha\beta\mu}P^{\alpha\beta}{}_\nu\right) = -8\pi T_{\mu\nu} \,,\label{efe}
\end{eqnarray}
where for notational simplicity, we write $f_Q=f^{\prime }(Q)$ and the
energy-momentum tensor $T_{\mu\nu}$ is given by 
\begin{eqnarray}
T_{\mu \nu } &=&-\frac{2}{\sqrt{-g}}\frac{\delta \sqrt{-g}\,{{\cal L}}_{m}}{\delta
g^{\mu \nu }}\,. \label{emt}
\end{eqnarray}
and varying (\ref{qqm}) with respect to the connection, one obtains
\begin{equation}
\nabla_\mu\nabla_\nu \left(\sqrt{-g} f_Q P^{\mu\nu}{}_\alpha \right) = 0\,.  \label{cfe}
\end{equation}

With the formalism of $f(Q)$ gravity specified, the conservation of the energy momentum tensor is ensured by the field equations.
In this discussion our main interest is to formulate the  gravitational field equations governing static and spherically symmetric spacetimes of (\ref{efe}) to the study of wormhole geometries.


\section{The wormhole geometry and the field equations} \label{sec3}
Consider the static spherically symmetric line element representing a  wormhole geometry is given by \cite{Morris:1988cz}
\begin{equation}\label{eq13}
ds^2= e^{\Phi(r)}dt^2-\frac{dr^2}{1-\frac{b(r)}{r}}-r^2(d\theta^2+\sin^2\theta d\phi^2),
\end{equation}
where $\Phi(r)$ and $b(r)$ are defined as the redshift and the shape functions, respectively. The radial coordinate $r$ is non-monotonic in the sense that it decreases from infinity to a
minimum value $b(r_0)= r_0$ and then it increases from $r_0$ back to infinity. The minimum value of the surface
 area is called the throat of the wormhole with $4\pi r^2$.
 Moreover, flaring out condition is one of the most fundamental property of the wormhole throat, which satisfy
the condition $\frac{b(r)-rb^{\prime}(r)}{b^2(r)}>0$ \cite{Morris:1988cz}, and at the throat  $b^{\prime}(r_0) < 1$ is also imposed. Another condition that needs to be satisfied is $1-b(r)/r > 0$.  Beside the above conditions, wormhole geometries 
have no horizons to maintain the criteria for traversability, which implies that $\Phi(r)$ must be finite everywhere.

The stress tensor for an anisotropic fluid compatible with spherical symmetry is
\begin{equation}\label{eq7}
T_{\mu\nu}=(\rho+P)u_\mu u_\nu- P_{\perp} g_{\mu\nu}+ (P-P_{\perp})\chi_{\mu}\chi_{\nu},
\end{equation}
which is mostly used for wormhole matter for consideration. 
Here, $\rho$ is the energy density,  $P$ the radial pressure and $P_{\perp}$ the tangential pressure, respectively. In the above equation $u^{\mu}$ represents the 4-velocity of the fluid, while $\chi_{\mu}$ is a spacelike vector along the direction of anisotropy. In Einstein gravity, the wormhole solutions are sustained by exotic matter sources
involving a stress-energy tensor that violates the null energy condition (NEC) (in fact, it violates all the energy conditions \cite{Visser:1995}). Note that the NEC asserts  $T_{\mu\nu}k^{\mu}k^{\nu} \geq 0$ 
for any null vector $k^{\mu}$. In the case of a stress-energy tensor of the form (\ref{eq7}), we have $\rho+P_{i} \geq 0$.

Following the discussion in Ref. \cite{Lin:2021uqa} (see Eq. (36)) the non-metricity scalar $Q$ for 
spherically symmetric configuration (\ref{eq13}) is given by
\begin{equation}
    \label{Qsc}
  Q= -\frac{b}{r^2}\left[ \frac{rb'-b}{r(r-b)}+\Phi'\right].
\end{equation}
In summary, inserting the metric (\ref{eq13}) and the anisotropic matter distribution (\ref{eq7}), into the equations of motion (\ref{efe}), we extract the nonzero components of the field equations \cite{Lin:2021uqa}
\begin{widetext}
\begin{eqnarray} 
8 \pi \rho(r) &=&  \frac{1}{2r^2}\left(1-\frac{b}{r}\right)\left[2 r f_{QQ} Q' \frac{b}{r-b}+f_Q\left(\frac{b}{r-b}(2+r\Phi')+ \frac{(2r-b)(b'r - b)}{(r-b)^2}\right)+f \frac{r^3}{r-b} \right], \label{fe1}\\
8 \pi P(r) &=& -\frac{1}{2r^2}\left(1-\frac{b}{r}\right) \left[2 r f_{QQ} Q' \frac{b}{r-b}+f_Q\left(\frac{b}{r-b}\left(2+ \frac{rb'-b}{r-b}+r\Phi'\right)-2r\Phi'\right)+f \frac{r^3}{r-b} \right], \label{fe2}\\
8 \pi P_{\perp}(r) &=& -\frac{1}{4r}\left(1-\frac{b}{r}\right)\left[-2 r \Phi' f_{QQ} Q' +f_Q\left(2\Phi' \frac{2b-r}{r-b}- r(\Phi')^2+\frac{rb'-b}{r(r-b)}\left(\frac{2r}{r-b}+r\Phi' \right)-2r\Phi''\right) \right.  \nonumber\\
&& \left. + 2f \frac{r^2}{r-b} \right], \label{fe3}
\end{eqnarray}
\end{widetext}
where $f \equiv f(Q)$, $f_{QQ}= \frac{d^2f(Q)}{dQ^2}$ and $f_{Q}= \frac{df(Q)}{dQ}$. Finally, we have three independent equations (\ref{fe1})-(\ref{fe3}) for our six unknown quantities, i.e.,
$\rho(r)$, $P(r)$, $P_{\perp}(r)$, $\Phi(r)$, $b(r)$ and $f(Q)$. Thus the above system of equations is under-determined, and it is possible to adopt  different strategies to construct wormhole solutions.  Here, we will focus on a particularly interesting case that follows a constant redshift function, $\Phi'$ = 0. With this assumption one can simplify the calculations considerably and provide interesting exact
wormhole solutions.     

\subsection{Specific case: $f(Q)= Q+ \alpha Q^2$}

Here, we consider a power-law form of function $f(Q)$ given by 
$f(Q)= Q+ \alpha Q^2$, where $\alpha$ is a constant. This model has been used for stellar structure with polytropic EoS \cite{Lin:2021uqa}. 

\subsubsection{Form function: $b(r) = r_0^2/r$}

Considering the specific choice for the form function $b(r) = r_0^2/r$ \cite{Lobo:2008zu},
the field equations, eqs. (\ref{fe1})-(\ref{fe3}), reduce to
\begin{eqnarray}
         \rho(r) &=& \frac{\text{r}_0^2}{ 8 \pi r^8} \left(\frac{2 \alpha  \text{r}_0^4 \left(9 \text{r}_0^2-14 r^2\right)}{\left(r^2-\text{r}_0^2\right)^2}-r^4\right),\\
         P(r) &=& -\frac{\text{r}_0^2}{ 8 \pi r^8} \left(\frac{10 \alpha  \text{r}_0^4 \left(\text{r}_0^2-2 r^2\right)}{\left(r^2-\text{r}_0^2\right)^2}+r^4\right),\\
        P_{\perp}(r)&=& \frac{\text{r}_0^2}{  8 \pi r^8} \left(r^4-\frac{2 \alpha  \text{r}_0^4 \left(\text{r}_0^2-2 r^2\right)}{\left(r^2-\text{r}_0^2\right)^2}\right).
\end{eqnarray}
The above components help us to determine the null energy condition (NEC) along the radial and tangential direction, which are
\begin{eqnarray}
\rho+P &=& -\frac{r^6 \text{r}_0^2-r^4 \text{r}_0^4+4 \alpha  \text{r}_0^6}{  4\pi (r^{10}-  r^8 \text{r}_0^2)}, \label{NE0}\\
\rho+P_{\perp}&=& \frac{2 \alpha  \text{r}_0^8-3 \alpha  r^2 \text{r}_0^6}{\pi  r^8 \left(r^2-\text{r}_0^2\right)^2}.\label{NE00}
\end{eqnarray}
For this specific case, we see that at the throat of the wormhole i.e., at $r=r_0$ the NEC along the radial and tangential directions become undefined. This shows that wormhole solution could not exists with this form function. Moreover, we have tried with other form functions like $b(r) = r_0$,  $b(r)= r_0+\gamma r_0 \left(1-\frac{r_0}{r}\right)$ and $b(r) = r e^{r_0-r}$, but all attempts go into vain.  Thus, we conclude that postulating a power-law form $f(Q)= Q+ \alpha Q^2$ is not suitable for wormhole solution. In next two sections, alternately, we suppose an inverse power-law model for $f(Q)$ gravity. We now proceed to the investigation of the physical implications of a non-trivial $f(Q)$-ansatz, studying the
possible existence of wormhole geometries supported by $f(Q)$ gravity theory. Such choices have widely been considered in $f(T)$ gravity, see Ref. \cite{Cai:2015emx,Myrzakulov:2010vz,Rani:2016zbd} for a discussion. But, other choices of form function are also possible, which we leave for further study.

\subsection{Specific solutions: $f(Q)= Q+ \frac{\alpha} {Q}$ }

\subsubsection{Form function: $b(r) = r_0^2/r$}
Considering the specific case of $f(Q)$ gravity i.e., $f(Q)= Q+ \frac{\alpha} {Q}$.  The stress-energy tensor profile for this specific case is given by
\begin{eqnarray}
        \rho(r) &=& \frac{\alpha  r^8 \left(20 r^2 \text{r}_0^2-11 r^4-9 \text{r}_0^4\right)-4 \text{r}_0^8}{32 \pi  r^4 \text{r}_0^6},\\
         P(r) &=& \frac{\alpha  r^8 \left(-24 r^2 \text{r}_0^2+13 r^4+11 \text{r}_0^4\right)-4 \text{r}_0^8}{32 \pi  r^4 \text{r}_0^6},\\
         P_{\perp}(r)&=& \frac{\alpha  r^8 \left(\text{r}_0^4-r^4\right)+4 \text{r}_0^8}{32 \pi  r^4 \text{r}_0^6}.
\end{eqnarray}
For the case of the NEC along the radial and tangential direction is provided by  
\begin{eqnarray}
        \rho+P &=& \frac{\alpha  r^8 \left(r^2-\text{r}_0^2\right)^2-4 \text{r}_0^8}{16 \pi  r^4 \text{r}_0^6}, \label{NE2}\\
         \rho+P_{\perp}  &=& -\frac{\alpha  r^4 \left(-5 r^2 \text{r}_0^2+3 r^4+2 \text{r}_0^4\right)}{8 \pi  \text{r}_0^6}. \label{NE22}
\end{eqnarray}

\begin{figure*}
    \includegraphics[scale=.4]{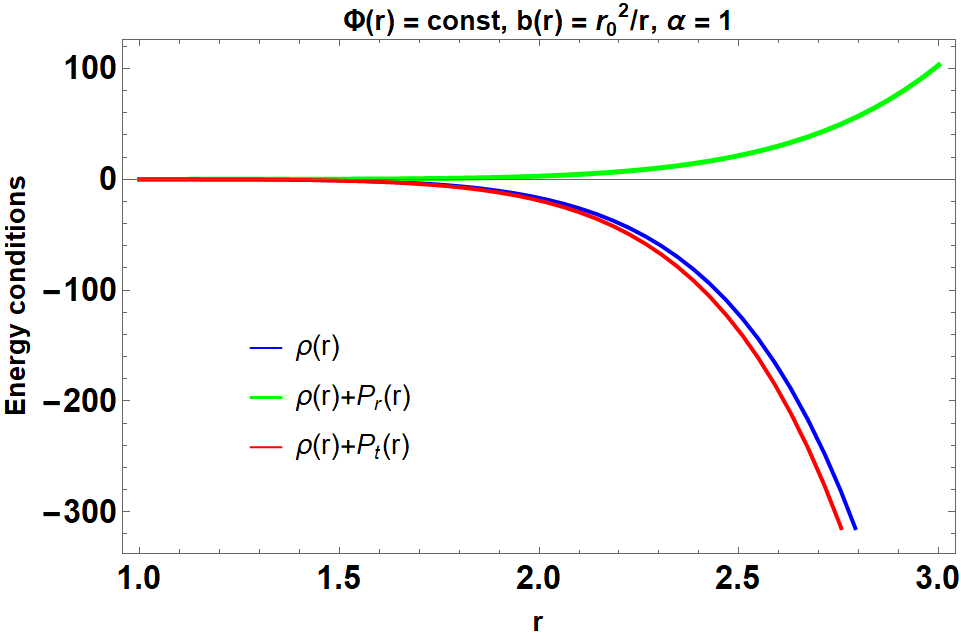}
    \includegraphics[scale=.4]{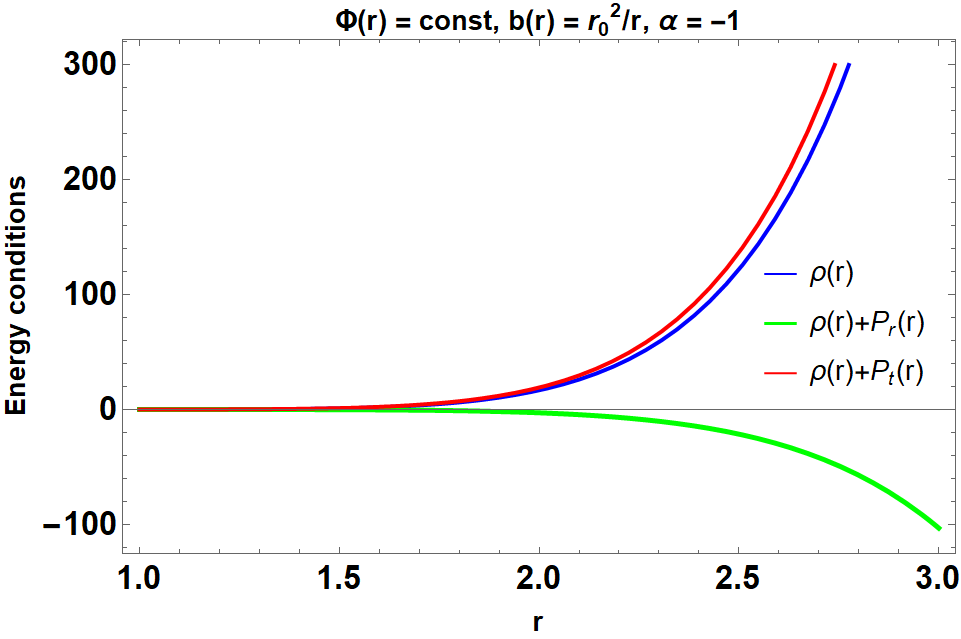} 
\caption{The figure represents the energy density and the null energy condition (NEC) for radial and tangential directions for the specific case of $f(Q)= Q+ \frac{\alpha} {Q}$, $\Phi'(r)=0$ and $b(r) = r_0^2/r$. We present the graphical behavior for $r_0=1$ and $\alpha =\pm 1$. The NEC given in Eqs. (\ref{NE2}) and (\ref{NE22}) is violated at the throat $r= r_0$ irrespective  of $\alpha$. }\label{f1}
\end{figure*}
 
 For concreteness, we plot graphs for energy density ($\rho$), $\rho+P$ and $\rho+P_{\perp}$ which are interpreted 
as the NEC along the radial and tangential direction, respectively. In Fig. \ref{f1}, we take into account the specific values
for $r_0 = 1$ and considered both cases of $\alpha = \pm 1$. It is interesting to observe that for $\alpha = 1$, the
energy density is positive whereas the NEC is violated throughout the spacetime. But, these situations are reversed
when we consider $\alpha = -1$, see right panel of Fig. \ref{f1}.

Moreover, one immediately finds from Eqs. (\ref{NE2}) and (\ref{NE22}) that $(\rho+ P)|_{r_0} = -\frac{1}{4 \pi  \text{r}_0^2}< 0$ and 
$(\rho+P_{\perp})|_{r_0} = 0$ at the throat or at its neighbourhood. This implies the violation of NEC for the normal matter threading
the throat of the wormhole. 

\subsubsection{Form function: $b(r) = r e^{r_0-r}$}
Here, we turn our attention to the model with $b(r) = r e^{r_0-r}$ \cite{Samanta:2018hbw}, where $0 < r_0< 1$ is particularly interesting to have wormhole solutions that satisfy the condition $b'(r_0)<1$. With this shape function the stress-energy tensor profile is given by
\begin{eqnarray}
        \rho(r) &=& \frac{e^{-r-3 \text{r}_0}}{8 \pi  r^2} \left[-3 \alpha  e^{4 r} (r+1) r^2-\alpha  (2 r+3) r^2 e^{2 (r+\text{r}_0)} \right. \nonumber\\ && \left. +\alpha  (5 r+6) r^2 e^{3 r+\text{r}_0}-(r-1) e^{4 \text{r}_0}\right],\\
         P(r) &=& \frac{e^{-r-3 \text{r}_0}}{8 \pi  r^2} \left[\alpha  e^{4 r} (4 r+3) r^2+3 \alpha  (r+1) r^2 e^{2 (r+\text{r}_0)} \right. \nonumber\\ && \left. -\alpha  (7 r+6) r^2 e^{3 r+\text{r}_0}-e^{4 \text{r}_0}\right],\\
         P_{\perp}(r)&=& \frac{1}{16 \pi  r}\left[\alpha  r^2 e^{r-\text{r}_0}-\alpha  r^2 e^{3 r-3 \text{r}_0}+e^{\text{r}_0-r}\right].
\end{eqnarray}

The NEC along the radial and tangential direction direction is given by 
\begin{figure*}
    \includegraphics[scale=.4]{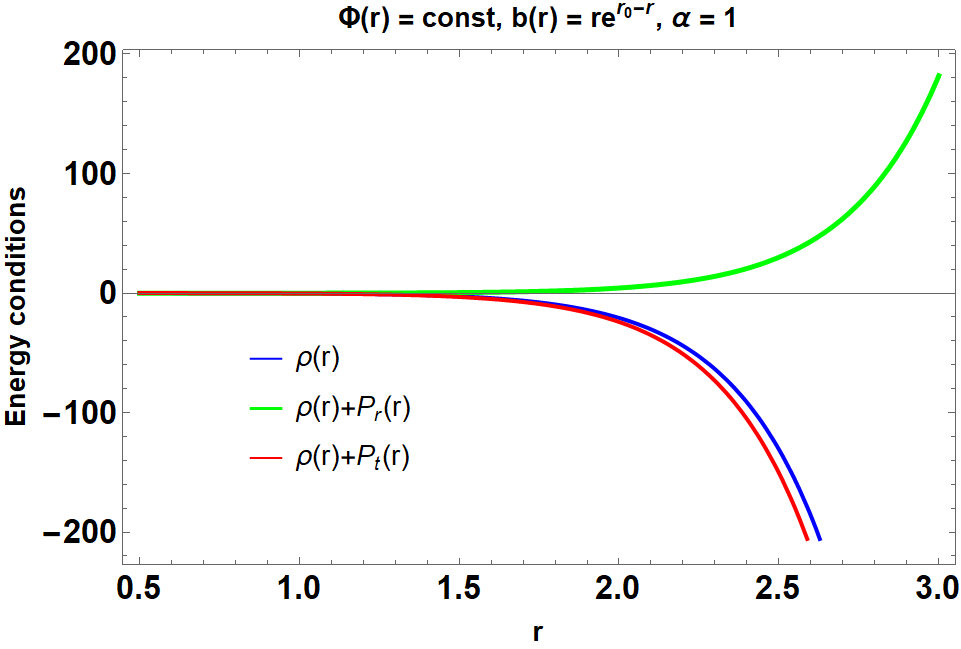}
    \includegraphics[scale=.4]{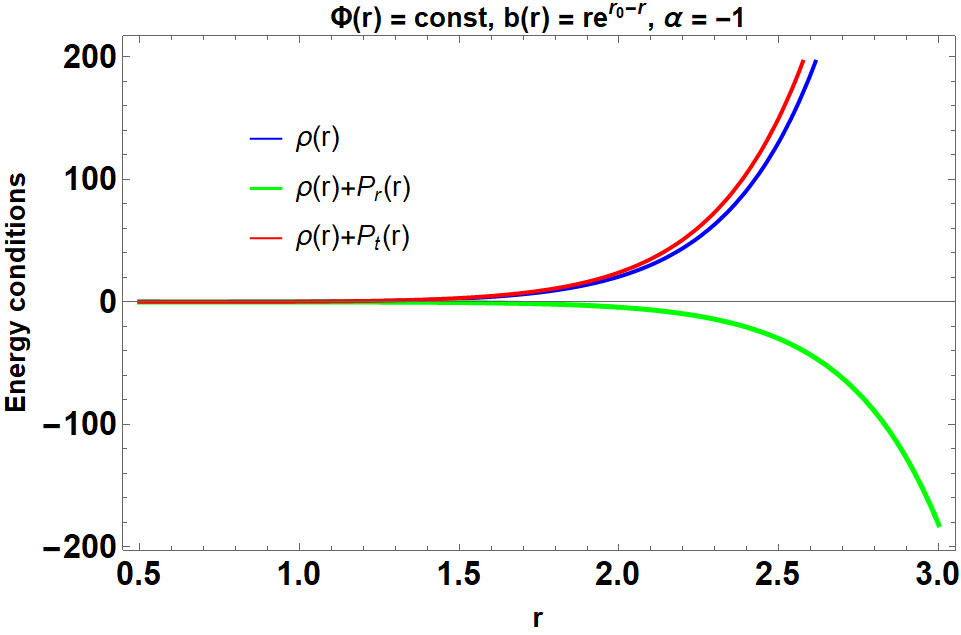} 
\caption{The graphical behavior of NEC in terms of $\rho$, $P$ and $P_{\perp}$ for the specific case of $f(Q)= Q+ \frac{\alpha} {Q}$, $\Phi'(r)=0$ and $b(r) = r e^{r_0-r}$.  The throat of the wormhole occurs at $r_0= 0.5$. For this case the standard NEC is always violated at the throat of the wormhole, see Eqs. (\ref{NE3}) and (\ref{NE33}). }\label{f2}
\end{figure*}

\begin{eqnarray}
        \rho+P &=& \frac{e^{-r-3 \text{r}_0}}{8 \pi  r} \left[\alpha  e^{4 r} r^2+\alpha  r^2 e^{2 (r+\text{r}_0)} -2 \alpha  r^2 e^{3 r+\text{r}_0}\right. \nonumber\\ && \left. -e^{4 \text{r}_0}\right], \label{NE3}\\ 
         \rho+P_{\perp}  &=& \frac{e^{-r-3 \text{r}_0}}{16 \pi  r^2} \left[-\alpha  e^{4 r} (7 r+6) r^2-3 \alpha  (r+2) r^2 e^{2 (r+\text{r}_0)} \right. \nonumber\\ && \left. +2 \alpha  (5 r+6) r^2 e^{3 r+\text{r}_0}-(r-2) e^{4 \text{r}_0}\right] . \label{NE33}
\end{eqnarray}

For this particular wormhole model we consider the throat at $r_0 =0.5$ and $b'(r_0) =0.5 <1$.  The graphical behavior of the $\rho$, $\rho+P$ and $\rho+P_{\perp}$ are presented on the left and right
side of Fig. \ref{f2} for $\alpha =\pm 1$.  This situation is same as of Fig. \ref{f1}, where the NEC is violated for  $\alpha = 1$ and satisfied for $\alpha = -1$.

We can also see from Eqs. (\ref{NE3}) and (\ref{NE33}) that $(\rho+ P)|_{r_0} = -\frac{1}{8 \pi  \text{r}_0}< 0$ and  $(\rho+P_{\perp})|_{r_0} = -\frac{\text{r}_0-2}{16 \pi  \text{r}_0^2} >0$ at the throat. This choice indicates that 
NEC is always violated at the wormhole throat. 

\subsection{Specific solutions: $f(Q)= Q \exp \left(\frac{\alpha }{Q}\right)$ }

\subsubsection{Form function: $b(r)= r_0^2/r$}

Using the form function $b(r)= r_0^2/r$, we find the following stress energy tensor components
\begin{widetext}
\begin{eqnarray}
        \rho(r) &=& e^{\frac{\alpha  r^4 (r^2-\text{r}_0^2) }{2 \text{r}_0^4}}\left[\frac{ \left(\alpha ^2 r^8 \left(5 r^2 \text{r}_0^2-3 r^4-2 \text{r}_0^4\right)+\alpha  r^6 \text{r}_0^4-2 \text{r}_0^8\right)}{16 \pi  r^4 \text{r}_0^6}\right],\\
         P(r) &=& e^{\frac{\alpha  r^4 (r^2-\text{r}_0^2)}{2 \text{r}_0^4}}\left[\frac{ \left(\alpha ^2 r^8 \left(-5 r^2 \text{r}_0^2+3 r^4+2 \text{r}_0^4\right)+\alpha  r^4 \text{r}_0^4 \left(r^2-2 \text{r}_0^2\right)-2 \text{r}_0^8\right)}{16 \pi  r^4 \text{r}_0^6}\right],\\
         P_{\perp}(r)&=& -e^{\frac{\alpha  r^4 (r^2-\text{r}_0^2) }{2 \text{r}_0^4}}\left[\frac{ \left(\alpha  r^6-2 \text{r}_0^4\right)}{16 \pi  r^4 \text{r}_0^2}\right].
\end{eqnarray}
\end{widetext}
The NEC along the radial and tangential direction are given by 
\begin{figure*}
    \includegraphics[scale=.4]{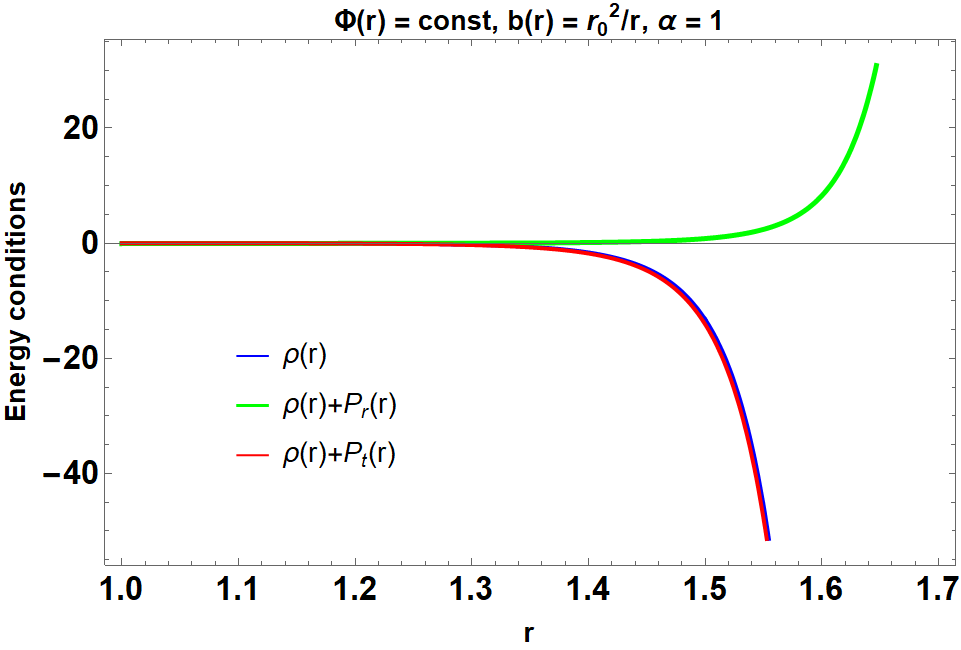}
    \includegraphics[scale=.4]{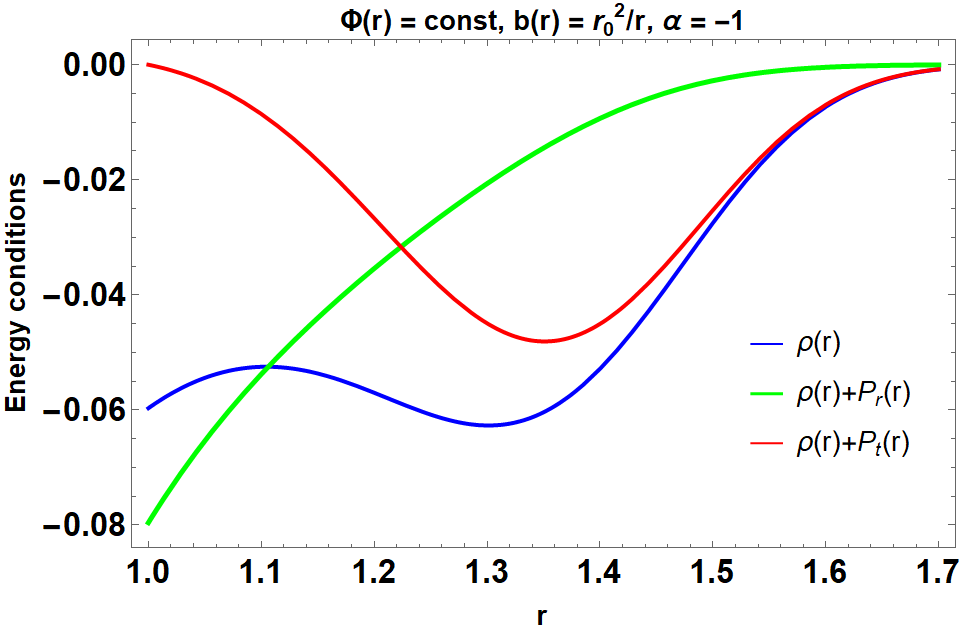} 
\caption{With $r_0= 1$ and $\alpha =\pm 1$, we plot $\rho$, $\rho+P$ and $\rho+P_{\perp}$ for the specific case of
$f(Q)= Q \exp \left(\frac{\alpha }{Q}\right)$, $\Phi'(r)=0$ and $b(r) = r_0^2/r$. The NEC is violated where the throat is located, see Eqs. (\ref{NE4}) and (\ref{NE44}). }\label{f3}
\end{figure*}

\begin{eqnarray}
        \rho+P &=& e^{\frac{\alpha  r^4 (r^2-\text{r}_0^2)}{2 \text{r}_0^4}}\left[\frac{ \left(\alpha  r^6-\alpha  r^4 \text{r}_0^2-2 \text{r}_0^4\right)}{8 \pi  r^4 \text{r}_0^2} \right], \label{NE4}\\
         \rho+P_{\perp}  &=&  e^{\frac{\alpha  r^4 (r^2-\text{r}_0^2) }{2 \text{r}_0^4}} \left[\frac{\alpha ^2 r^4 \left(5 r^2 \text{r}_0^2-3 r^4-2 \text{r}_0^4\right) }{16 \pi  \text{r}_0^6}\right].\label{NE44}
\end{eqnarray}

From the graphical behavior of the NEC in terms of $\rho+P$ and $\rho+P_{\perp}$, presented in Fig. \ref{f3}, we see that 
NEC is always violated for $\alpha= \pm 1$. 

Also, we can also see from Eqs. (\ref{NE4}) and (\ref{NE44}) that $(\rho+ P)|_{r_0} = -\frac{1}{4 \pi  \text{r}_0^2}< 0$ and  $(\rho+P_{\perp})|_{r_0} = 0$, and thus the standard NEC becomes violated at the close vicinity of the wormhole throat.



\section{Concluding remarks}\label{sec8}
Wormholes are hypothetical objects connecting two asymptotic regions or infinities,  possibly through which observers may freely traverse. But the main challenge in wormhole physics is to find a matter source without violating the energy conditions. Recently, in \cite{Lin:2021uqa}, authors have investigated the external and internal solutions of spherically symmetric objects in $f(Q)$ gravity.  Interesting the vacuum solution obtained in \cite{Lin:2021uqa} for $f(Q)$ is exactly same as reported in \cite{DAmbrosio:2021zpm}. Following this approach, we have explored wormhole geometries in the 
framework of $f(Q)$ gravity for static and spherically symmetric spacetime. More accurately, we focused the analysis based on the specific choices for the $f(Q)$ form and shape functions. We simplify our calculations by assuming constant redshift function i.e., $\Phi'=0$ and to avoid the presence of event horizons.

The first attempt is a phenomenological power law $f(Q)= Q+ \alpha Q^2$, where we found that wormhole solutions could not exist because the energy density and two pressure components are in indeterminate forms at the throat. The next two attempts base on the inverse power law of
$f(Q)= Q+ \frac{\alpha} {Q}$ and $f(Q)= Q \exp \left(\frac{\alpha }{Q}\right)$, respectively.  By carefully considering a  specific shape function, we solved the field equations for $f(Q)$ gravity and obtained energy density and pressure profiles that needed to support the wormhole geometries. In every case we have found a similar situation for $\alpha=  1$,  where the energy density is positive with violation of NEC throughout the spacetime.  For $\alpha=  -1$, we found negative  energy density but obeying the NEC extending outward from the throat. However, in any case of $\alpha=  \pm 1$, one verifies that the NEC is violated at the throat of the wormhole. 

Our findings are completely different with the solution reported in \cite{Mustafa:2021ykn}. In \cite{Mustafa:2021ykn}, authors have shown the possibility of obtaining traversable wormholes satisfying the energy conditions using Karmarkar conditions with embedded class-1 spacetime. In our case we have studied a wide variety of exact solutions of asymptotically flat spacetimes, but all solutions violate the NEC at the throat of the wormhole.


\section*{Acknowledgments}
The author TT would like to thank the financial support from the Science Achievement Scholarship of Thailand (SAST). A. Pradhan thanks to IUCCA, Pune, India for providing facilities under associateship programmes.

\end{document}